\documentclass[a4paper,12pt]{article}
\usepackage[brazil]{babel}
\usepackage[latin1]{inputenc}

\usepackage{epsfig}
\usepackage{amsmath}
\usepackage{amstext,mathrsfs}
\usepackage{amsfonts}
\usepackage{amssymb}

\def\beq{\begin{equation}}
\def\eeq{\end{equation}}
\def\beqa{\begin{eqnarray}}
\def\eeqa{\end{eqnarray}}
\def\ban{\begin{eqnarray*}}
\def\ean{\end{eqnarray*}}
\def\barr{\begin{array}}
\def\earr{\end{array}}
\def\n{\noindent}
\def\m{\mid}
\def\la{\langle}
\def\ra{\rangle}
\def\s{\sum}
\def\h{\hbar}

\begin{document}

\thispagestyle{empty}

\begin{center}

{\Large \bf Dissipative Quantum Systems in ThermoField
Dynamics}\footnote{Work presented in the Workshop on Quantum
Nonstationary Systems, held in the International Centre for
Condensed Matter Physics at the University of Brasilia, from October
19th-23rd.}

\bigskip

\emph{J.L. Tomazelli}$^{\dag}$ and \emph{G.G. Gomes}$^{\ddag}$

\medskip

Departamento de Física - Universidade Federal de Santa Catarina

\medskip

{\scriptsize $^{\dag}$tomazelli@fsc.ufsc.br,
$^{\ddag}$gerson.g.gomes@gmail.com}

\end{center}

\medskip

{\footnotesize We investigate a class of microscopic systems in
interaction with a macroscopic system in thermal equilibrium,
following the construction of Dalibard, Dupont-Roc and
Cohen-Tannoudji (DDC). By considering self-adjoint operators as
elements of Schwinger's Measurement Algebra (SMA), we construct
statistical mean values of the relevant observables as matrix
elements in a suitable operator basis, which correspond to the
vacuum states of ThermoField Dynamics (TFD).}

\vspace{0.8cm}

\n{\large \bf 1. Schwinger Operadors and TFD}

\bigskip

In SMA [1] an operator is defined as

\begin{equation}
X = \s_{a',a''} \la a' \m X \m a'' \ra M(a',a'')\, ,
\end{equation}
where $\la a' \m X \m a'' \ra$ represents its matrix element in the
$a$ representation:

\beq \la a' \m X \m a'' \ra = Tr X M(a'',a').\eeq

\n The product of two general measurement symbols, referring to
distinct sets of compatible observables $A,B,C$ and $D$, satisfies
the following composition rule: \beq M(a',b') M(c',d') = \la
b'|c'\ra M(c',d'),\eeq

\n where the number $\la b'|c'\ra$ defines the statistical relation
between the corresponding sets.

\n The expectation value of a given property $A$ in the $b'$ basis
is \beq \la A \ra_{b'} = \s_{a'} a' p(a',b') = Tr A M(b') = \la b'
\m A \m b' \ra \, , \eeq

\n where we introduce the operator
\beq A = \s_{a'} a' M(a') =
\s_{a',a''} a' \delta(a',a'') M(a',a'') \, , \eeq

\n in the basis $M(a') \equiv M(a',a')$. This implies that \n $\la
a' \m A \m a''\ra = a' \delta(a',a'')$.

\n Now we define the statistical mean

\beq \la A \ra = \s_{b'} \Pi(b') \la A \ra_{b'} = \s_{a'} \la a' \m
\rho A \m a' \ra = Tr (\rho A) ,\eeq

\n with $\Pi(b') \geq 0$ and $\s_{b'} \Pi(b') = 1$, where
\beq \rho
\equiv \s_{b'} \Pi(b') M(b') = \s_{b',b''} \Pi(b') \delta(b',b'')
M(b',b'')\, . \eeq

\n Hence $\la b' \m \rho \m b'' \ra = \Pi(b') \delta(b',b'') \, , $
with $Tr \rho = \s_{b'} \Pi(b') = 1$. In thermal equilibrium, by
considering the basis $M(n,m)$ associated to the number operator $N$
in which its matrix elements are diagonal, \beqa N &=& \s_{n,m} \la
n \m N \m m \ra \delta_{nm} M(n,m) \nonumber
\\ &=& \s_{n} \la n \m N \m n \ra M(n) \, ,\eeqa

\n we have
\beq \rho = \s_{n} \Pi(E_n) M(n) \, , \eeq

\n where
\beq \Pi(E_n) = Z^{-1} e^{- \beta E_n} \, , \eeq \n

\n or \beq \rho = \s_{n,m} \Pi(E_n) \delta_{n m} M(n,m) \, . \eeq

\n Introducing an auxiliary operator basis where a new fictitious
operator $\tilde{N}$, corresponding to $N$, is diagonal, we can
write $\delta_{n m} M(n,m)$ as

\beq  \la \tilde{n} \m \tilde{m} \ra M(n,m) = M(n,\tilde{n})
M(\tilde{m},n)\, , \eeq

\n so that
\beqa \rho &=& \s_{n,m} \sqrt{\Pi(E_n) \Pi(E_m)}
M(n,\tilde{n}) M(\tilde{m},n) \nonumber \\&=& \left[ \s_{n}
\sqrt{\Pi(E_n)} M(n,\tilde{n}) \right] \left[ \s_{m} \sqrt{\Pi(E_m)}
M(\tilde{m},m)
\right] \nonumber \\
&\equiv& \left| 0(\beta) \ra \la 0(\beta) \right| \, . \eeqa

\n We may identify the measurement symbols inside the square
brackets as composite states of the thermal vacuum of TFD [2]: \beq
\m 0(\beta) \ra \equiv \s_n \sqrt{\Pi(E_n)} \m n, \tilde{n} \ra \,
,\label{estadoTFD}\eeq

\n so that $\rho$ acquires the status of a projection operator, as
well as $M(n,\tilde{n}) \equiv |\tilde{n}\ra \la n |$ and
$M(\tilde{m},m) \equiv |\tilde{m}\ra \la m |$. Therefore, for a
given observable $F$, \beqa Tr \left(\rho F \right) &=& \la 0(\beta)
\m F \m 0(\beta) \ra = Tr \left( F \m 0(\beta) \ra \la 0(\beta) \m \right) =\nonumber \\
&=& Tr \left(\s_{n,m} \sqrt{\Pi(E_n) \Pi(E_m)} \, |\tilde{m} \ra \la
m | F | n \ra \la \tilde{n} | \right) \nonumber \\
&=& \s_{n,m} \sqrt{\Pi(E_n) \Pi(E_m)} \, \la
\tilde{n}|\tilde{m} \ra \la m | F | n \ra \nonumber \\
&=& \s_n \Pi(E_n) \la n | F | n \ra \, .\label{vacuoTFD}\eeqa

\medskip


\n{\large \bf 2. Radiation Considered as a Reservoir}

\bigskip

Let us consider the problem of a small system $\mathcal{A}$
interacting with a large reservoir $\mathcal{R}$. Following DDC
construction [3], let \beq H = H_A + H_R + V ,\label{hamiltotal}\eeq

\n be the Hamiltonian of global system $\mathcal{A} + \mathcal{R}$,
where $H_A$ and $H_R$ are, respectively, the Hamiltonians of
$\mathcal{A}$ and $\mathcal{R}$, whilst $V$ stands for the
interaction between them. In the interaction picture, the density
operator of the global system obeys the evolution equation, \beq
\frac{d}{dt} \rho(t) = \frac{1}{i \hbar} [V(t), \rho(t)] ,
\label{evolint} \eeq

\n from which we obtain \beqa \rho(t + \Delta t) & = & \rho(t) +
\frac{1}{i \hbar} \int_t^{t + \Delta t}[V(t'),\rho(t)] dt' +
\nonumber\\ & &+ \left(\frac{1}{i \hbar} \right)^2 \int_t^{t +
\Delta t} dt' \int_t^{t'} [V(t'),[V(t''),\rho(t'')]] dt'' .
\label{B6} \eeqa

\n By taking the trace with respect to $\mathcal{R}$, we arrive at
\beqa \Delta \sigma(t) &\equiv& \sigma(t + \Delta t) - \sigma(t) =
\frac{1}{i \hbar} \int_t^{t +
\Delta t} Tr_R[V(t'),\rho(t)] dt' \; + \nonumber\\
&+& \left(\frac{1}{i \hbar} \right)^2 \int_t^{t + \Delta t} dt'
\int_t^{t'} Tr_R[V(t'),[V(t"),\rho(t'')]] dt'' \, , \label{B9} \eeqa

\n where
\beq \sigma(t) = Tr_R \rho(t) .\label{opdensR}\eeq

\n The interaction $V$ between $\mathcal{A}$ e $\mathcal{R}$ will be
taken as a product of an observable $A$ of $\mathcal{A}$ and an
observable $R$ of $\mathcal{R}$:

\beq V = - A R .\label{acoplint}\eeq

\n Since the average value in $\sigma_R$ of the coupling $V(t)$ is
zero, the leading contribution in (\ref{B9}) stems from the two-time
average \beq g(t',t'') = Tr_R[\sigma_R R(t') R(t'')]
.\label{gdtau}\eeq

\n If $V$ is sufficiently small, and $\Delta t$ sufficiently short
compared with the evolution time $t_R$ of $\sigma$, $\rho(t)$ can be
written in the form \beq \rho(t) = Tr_R \rho(t) \otimes Tr_A
\rho(t),\eeq

\n where the contributions of the correlations between $\mathcal{A}$
and $\mathcal{R}$ in the time $t$ were neglected. The general idea
is that the initial correlations between $\mathcal{A}$ and
$\mathcal{R}$ at time $t$ disappear after a collision time $\tau_c
\ll t_R$. Thus, there exist two very different time scales, such
that \beq \tau_c \ll \Delta t \ll t_R, \label{escalast} \eeq

\n bringing us to the 'coarse-grained' rate of variation for the
system $\mathcal{A}$.

\bigskip


\n{\large \bf 3. Master Equation for a Damped Harmonic Oscillator}

\bigskip

\n{\bf 3.1. The Physical System}

\bigskip

We are interested in the case where the small system $\mathcal{A}$
is a one-dimensional harmonic oscillator of frequency $\omega_0$
whose Hamiltonian is \beq H_A = \hbar \omega_0 (b^\dag b +
\frac{1}{2}) , \label{hamiltA}\eeq

\n where $b^\dag$ and $b$ are the rising and lowering operators of
this oscillator. The reservoir $\mathcal{R}$ consists of an infinite
number of one-dimensional harmonic oscillators, of frequency
$\omega_i$, with ladder operators $a_i^\dag$ and $a_i$, so that the
Hamiltonian $H_R$ for $\mathcal{R}$ is written as \beq H_R = \s_i
\hbar \omega_i (a_i^\dag a + \frac{1}{2}) . \label{hamiltR}\eeq

\n We take a sesquilinear interaction between $\mathcal{A}$ and
$\mathcal{R}$ of the form \beq V = V^\dag = \s_i (\eta_i b^\dag a_i
+ \eta_i^* b a_i^\dag) , \label{sesq}\eeq

\n where $\eta_i$ is the coupling constant between $\mathcal{A}$ and
the $i$-th oscillator  of $\mathcal{R}$.

\bigskip


\n{\bf 3.2. The Master Equation}

\bigskip

We write the coarse-grained rate of variation,
\beq \frac{\Delta
\sigma}{\Delta t} = - \frac{1}{\hbar^2} \frac{1}{\Delta t}
\int_0^{\infty} d(t'-t'') \int_t^{t+\Delta t}dt' Tr_R
[V(t'),[V(t''),\sigma_A(t) \otimes \sigma_R]] ,
\label{coarsegrained2}\eeq

\n with $V$ given by (\ref{sesq}). Changing to the Schrodinger
representation, we obtain the following operator form for the master
equation: \beqa \frac{d \sigma}{dt} = &-&
\frac{\Gamma}{2}[\sigma,b^\dag b]_+ -
\Gamma'[\sigma,b^\dag b]_+ - \Gamma' \sigma \nonumber\\
&-& i (\omega_0 + \Delta) [b^\dag b,\sigma] + \Gamma b \sigma b^\dag
+ \Gamma'(b^\dag \sigma b + b \sigma b^\dag)
\,.\label{eqmestra}\eeqa

\n In this equation $[ \;, ]_+$ represents the anticommutator and we
have made the following definitions \beq \Gamma = \frac{2
\pi}{\hbar} \s_i \m \eta_i \m^2 \delta(\hbar \omega_0 - \omega_i)\,,
\quad \Gamma' = \frac{2 \pi}{\hbar} \s_i \m \eta_i \m^2 \la n_i \ra
\, \delta(\hbar \omega_0 - \omega_i) ,\label{Gamalinha} \eeq

\n and \beq \hbar \Delta = V.P.\left(\s_i \frac{\m \eta_i
\m^2}{\hbar \omega_0 - \hbar \omega_i} \right)\,, \quad  \hbar
\Delta' = V.P.\left(\s_i \frac{\m \eta_i \m^2 \la n_i \ra}{\hbar
\omega_0 - \hbar \omega_i} \right).\label{Deltalinha}\eeq

\n Here $\la n_i \ra$ is the average number of excitation quanta of
the oscillator $i$. If this number depends only on the energy of
this oscillator, due to delta function in the second equation in
(\ref{Gamalinha}), we have \beq \Gamma' = \la n(\omega_0) \ra \Gamma
, \label{eq15}\eeq

\n where $\la n(\omega_0) \ra$ is the average number of quanta in
the reservoir oscillators, having the same frequency $\omega_0$ of
oscillator $\mathcal{A}$. If, moreover, $\mathcal{R}$ is in
thermodynamic equilibrium, $\la n(\omega_0) \ra$ is equal to
$[\exp{\h \omega_0 / k_B T} - 1]^{-1}$.
\newline

\n Alternatively, temperature effects can be incorporated from the
very beginning, replacing the trace in (\ref{coarsegrained2}) by the
thermal expectation value (\ref{vacuoTFD}), leading to \beq
\frac{\Delta \sigma}{\Delta t} = - \frac{1}{\hbar^2} \frac{1}{\Delta
t} \int_0^{\infty} d(t'-t'') \int_t^{t+\Delta t}dt' \la 0(\beta) |
[V(t'),[V(t''),\sigma_A(t)]] | 0(\beta) \ra ,\label{coarsegrTFD}\eeq

\n and, consequently, to the master equation (\ref{eqmestra}). In
this case, the average number $\la n_i \ra$ in (\ref{Gamalinha}) and
(\ref{Deltalinha}) is given by a mean value in the TFD vacuum state,
defined in (\ref{estadoTFD}), \beq \la n_i \ra = \la
0(\beta)|a_i^\dag a_i | 0(\beta) \ra ,\eeq

\n after working out the thermal Green function \beqa g(\tau) &=& Tr
\left[ R(t) R(t-\tau) | 0(\beta)
\ra \la 0(\beta) | \right] \nonumber \\
&=& \la 0(\beta) | R(t) R(t - \tau) | 0(\beta) \ra ,\eeqa

\n instead of the two-time average (\ref{gdtau}), \n with $\tau
\equiv t' - t''$.
\newline

\n As a straightforward application of the formalism just presented,
equation (\ref{coarsegrTFD}) can be employed to derive the
corresponding master equation (\ref{eqmestra}), in order to study a
brownian particle interacting with a macroscopic system in the
framework of equilibrium TFD [4].

\bigskip

\n{\large \bf References}

\bigskip

\n [1] J.S. Schwinger, {\it Quantum Kinematics and Dynamics},
Addison-Wesley (1991).
\newline

\n [2] H. Umezawa and Y. Takahashi, Int. J. Mod. Phys. B {\bf 10}
(1996) 1755; H. Umezawa, {\it Advanced Field Theory} AIP Publishing
(1993).
\newline

\n [3] C. Cohen-Tannoudji, J. Dupont-Roc and G. Grynberg, {\it
Atom-Photon Interactions: Basic Processes and Applications}, Wiley
(2004).
\newline

\n [4] J.L. Tomazelli and G.G. Gomes, work in progress.

\end{document}